\documentclass[english,aip,pof,amsmath,amssymb,longbibliography,preprint]{revtex4-2}
\usepackage[T1]{fontenc}
\usepackage[latin9]{inputenc}
\usepackage{color}
\usepackage{amssymb}
\usepackage{graphicx}

\makeatletter

\providecommand{\tabularnewline}{\\}

\usepackage{subfigure}

\makeatother

\usepackage{babel}
\begin{document}
\title{Fluid mode spectroscopy for measuring kinematic viscosity of fluids
in open cylindrical containers}
\author{Hideshi Ishida}
\email{hideshi.ishida@mec.setsunan.ac.jp}

\author{Masaaki Horie}
\author{Takahiro Harada}
\author{Shingo Mizuno}
\author{Seita Hamada}
\author{Haruki Imura}
\author{Shoma Ashiwake}
\author{Naoya Isayama}
\author{Ryomei Saeki}
\author{Ryotaro Kozono}
\author{Daichi Taki}
\author{Asuka Kurose}
\affiliation{Department of Mechanical Engineering, Faculty of Science and Engineering,
Setsunan University, 17-8 Ikeda-Nakamachi, Neyagawa, Osaka 572-8508,
Japan}
\begin{abstract}
On a daily basis we stir tee or coffee with a spoon and leave it to
rest. We know empirically the larger the stickiness, viscosity, of
the fluid, more rapidly its velocity slows down. It is surprising,
therefore, that the variation, the decay rate of the velocity, has
not been utilized for measuring (kinematic) viscosity of fluids. This
study shows that a spectroscopy decomposing a velocity field into
fluid modes (Stokes eigenmodes) allows us to measure accurately the
kinematic viscosity. The method, Fluid Mode Spectroscopy (FMS), is
based on the fact that each Stokes eigenmode has its inherent decay
rate of eigenvalue and that the dimensionless rate of the slowest
decaying mode (SDM) is constant, dependent only on the normalized
shape of a fluid container, obtained analytically for some shapes
including cylindrical containers. The FMS supplements major conventional
measuring methods with each other, particularly useful for measuring
relatively low kinematic viscosity and for a direct measurement of
viscosity at zero shear rate without extrapolation. The method is
validated by the experiments of water poured into an open cylindrical
container, as well as by the corresponding numerical simulations. 
\end{abstract}
\maketitle

\section{Introduction}

Since Isaac Newton stated in 1687 that the resistance in the parts
of a fluid is proportional to the velocity with which the parts of
the fluid are separated from one another \citep{partal2010}, viscosity
has been an important physical quantity in fluid mechanics and engineering.
However, it took nearly 200 years for viscosity to be measured until
the dynamical equation of fluid was established\citep{navier1822,stokes1845}.

To our best knowledge, Hagenbach was the first in the world to measure
viscosity and report its value in an academic journal, Poggendorff's
Annalen, in 1860 \citep{hagenbach1860}. He measured the viscosity
of water with the Hagen-Poiseuille equation \citep{sutera1993} while
changing the water temperature. Later, Holman \citep{holman1877,holman1886}
measured the viscosity for various fluids and temperatures by 1886.
In 1894, Ostwald \citep{ostwald1894} established the method of capillary
viscometer, which became commonly used to measure the viscosity of
fluids \citep{flowers1914,kawada1955,gupta2014,capillary2020a,capillary2020b,fallingb2020,fallingb2021,rotational2019a,rotational2019b,osci2019,oscil2023}.

According to Flower \citep{flowers1914}, Kawada \citep{kawada1955}
and Gupta \citep{gupta2014}, there are four major methods for measuring
the viscosity of fluids, i.e. capillary \citep{capillary2020a,capillary2020b},
falling ball/piston \citep{fallingb2020,fallingb2021}, rotational
\citep{rotational2019a,rotational2019b}, and oscillating \citep{osci2019,oscil2023}
viscometers, and they have been already established by 1914. For example,
the time taken for the fluid to flow is measured by a capillary viscometer.
Similarly, the torque required to maintain a constant speed is measured
by a rotational viscometer, logarithmic decay rate by an oscillating
viscometer, and the time taken for an object to pass through by a
falling ball/piston viscometer. In total, these conventional methods
have a feature that the measured quantity increases with viscosity.
It follows that they tend to decrease its measurement accuracy and
to require larger equipment for its improvement when the viscosity
or shear stress is small, although the tendency does not necessarily
make the measurement of viscosity impossible. Furthermore, the viscosity
generally depends on physical properties such as shear rate and temperature,
which are not uniform inside the finite experimental apparatus. Consequently,
the measurement involves errors or the need for corrections based
on the condition of the experimental apparatus, posing the problem
of determining the temperature, pressure, and shear rate at which
the viscosity is measured.

In this study, a new method of FMS (Fluid Mode Spectroscopy) for measuring
kinematic viscosity of fluids is proposed based on a spectroscopy
that decompose a velocity field into fluid modes, Stokes eigenmodes
\citep{leriche2008,labrosse2014}, with inherent decay rates. The
method, regarded as one of mode spectroscopy methods for measuring
diffusion coefficients \citep{rus93,rus_ogi99,ogi16,ishida18}, measures
the (exponential) decay rate of fluid speed after the fluid is stirred,
evaluating the viscosity in nearly a spatially uniform, zero-shear-stress,
stationary state of the fluid. In principle, it is easy for us to
apply the method to fluids with low viscosity, such as water, without
corrections by prolonging the measurement time. Its accuracy depends
only on the used velocimetry. Inversely, its application to larger
viscosity fluids tends to diminish the accuracy because of the measurement
of decay rate within a shorter period. In addition, it cannot be utilized
as a rheometry, i.e. a viscometry for non-Newtonian fluids under non-zero
shear stresses at this time. Instead, direct measurement of the kinematic
viscosity at zero shear stress without extrapolation is possible for
FMS, while impossible for conventional methods. The proposed method
is useful in the sense that FMS and conventional methods complement
each other.

In recent years, new methods for the viscosity measurement have been
proposed. They include viscometries by using a quartz crystal resonator
\citep{QCR2021a}, ultrasonic shear-wave reflectance \citep{uswr2019a},
an ultrasonic transducer in a reserve tank \citep{utrt2022a}, droplet
microfluidics \citep{microdroplet2021a}, oscillating drops \citep{oscillating_drop2020a},
surface distortion caused by a pulsed gas jet \citep{gasjet2020a},
and others \citep{md2021a,penetration2018a,takeda2020development,freely_decaying_osci2018a,secondary_flow2022a}.
To our best knowledge, however, a spectroscopy decomposing a velocity
field into decaying fluid modes has not been utilized as a viscometer.
This study is the first trial to apply the method to measuring the
viscosity of water in a cylindrical container with a free, top surface,
and its applicability and accuracy are investigated.

\section{Fluid Mode Spectroscopy (FMS)}

\subsection{Fundamentals of FMS\label{subsec:Fund_FMS}}

The principle of Fluid Mode Spectroscopy (FMS) is so simple. After
a fluid in an open or closed container is stirred to give an initial
velocity field $\mathbf{v}^{*}$ of position $\mathbf{x}^{*}$ at
time $t^{*}=0$, the field decays to vanish without forcing. In this
study, asterisk $*$ denotes dimensional quantities. The transient
field can be typically expressed as the superposition of fluid modes
$\mathbf{v}_{1}^{*},\mathbf{v}_{2}^{*},\cdots$ as follows

\begin{equation}
\mathbf{v^{*}}(\mathbf{x}^{*},t^{*})=e^{-\lambda_{1}^{*}t^{*}}\mathbf{v}_{1}^{*}(\mathbf{x})+e^{-\lambda_{2}^{*}t^{*}}\mathbf{v}_{2}^{*}(\mathbf{x})+\cdots,\label{eq:transient_velocity}
\end{equation}
where $0<\lambda_{1}^{*}\leq\lambda_{2}^{*}\leq\cdots$. Such modes
are well known to be the Stokes eigenmodes \citep{leriche2008,labrosse2014}.
The (exponential) decay rate $\lambda_{i}^{*}$ is the real eigenvalue
corresponding to the $i$th eigenvector field $\mathbf{v}_{i}^{*}$,
and $\mathbf{v}_{1}^{*}$ is the slowest decaying mode (SDM). Since
the RHS of Eq. (\ref{eq:transient_velocity}) is dominated by the
SDM after sufficiently long time, we can eventually detect the decay
rate $\lambda_{1}^{*}$ of $\mathbf{v}^{*}$ as the gradient on a
semi-log plot. 

The Buckingham $\Pi$ theorem leads to that the physical quantities
in Eq. (\ref{eq:transient_velocity}) are normalized by the kinematic
viscosity $\nu$ of the fluid and typical length $L^{*}$ and that
the normalized decay rate $\lambda_{SDM}(\equiv\lambda_{1}^{*}L^{*^{2}}/\nu)$
of the SDM must depend only on the normalized shape of a fluid container.
The rate is obtained by solving a dimensionless eigen equation for
the Stokes eigenmodes. For the case of an axisymmetric, radius-component-free
flow in a horizontally positioned cylindrical container with a flat
upper, open surface, the rate is analytically found to be

\begin{equation}
\lambda_{SDM}=x_{1}^{2}+\left(\frac{\pi}{2\alpha}\right)^{2},\label{eq:lamda_SDM}
\end{equation}
where $x_{1}(\doteqdot3.83170597)$ denotes the positive, smallest
zero of the Bessel function of the first kind of order unity, $J_{1}(x)$,
and $\alpha$ the aspect ratio of depth $D^{*}$ to the inner radius
$R^{*}(=L^{*})$ of the container. The azimuthal velocity component
$u_{\theta}$ of its corresponding eigenvector field is of the form

\begin{equation}
u_{\theta}\sim J_{1}(x_{1}r)\sin\left(\frac{\pi z}{2\alpha}\right)\label{eq:Bessel_dist}
\end{equation}
where $r(=r^{*}/R^{*})$ denotes dimensionless radius, and $z(=z^{*}/R^{*})$
dimensionless height. See Appendix \ref{sec:Stokes-modes-cc}.

Thus, the kinematic viscosity of a fluid in a container can be evaluated
by

\begin{equation}
\nu=\lambda_{1}^{*}L^{*^{2}}/\lambda_{SDM},\label{eq:evald_viscosity}
\end{equation}
with a measured (dimensional) decay rate $\lambda_{1}^{*}$. This
is the essence of FMS.

Rigorously speaking, the expression (\ref{eq:transient_velocity})
makes physical sense for the case that all eigenvalues are semi-simple.
Since the negative gradient of $\ln(t^{b}\exp(-\lambda_{SDM}t))$
is well approximated by $\lambda_{SDM}$ for a constant $b$, even
if not so, the value of $\lambda_{SDM}$, whether obtained analytically
or numerically, makes it possible for us to utilize the method of
FMS. 

In order to avoid numerical errors, it is desirable that $\lambda_{SDM}$
is obtained analytically. Moreover, the area of flow visualization
can be limited to the top surface for measuring the decay rate $\lambda_{1}^{*}$
for the case of an open container. They are the reason why we conducted
experiments by use of an open cylindrical container described in the
next section.

\subsection{Experimental procedure}

\begin{figure}
\includegraphics[scale=0.5]{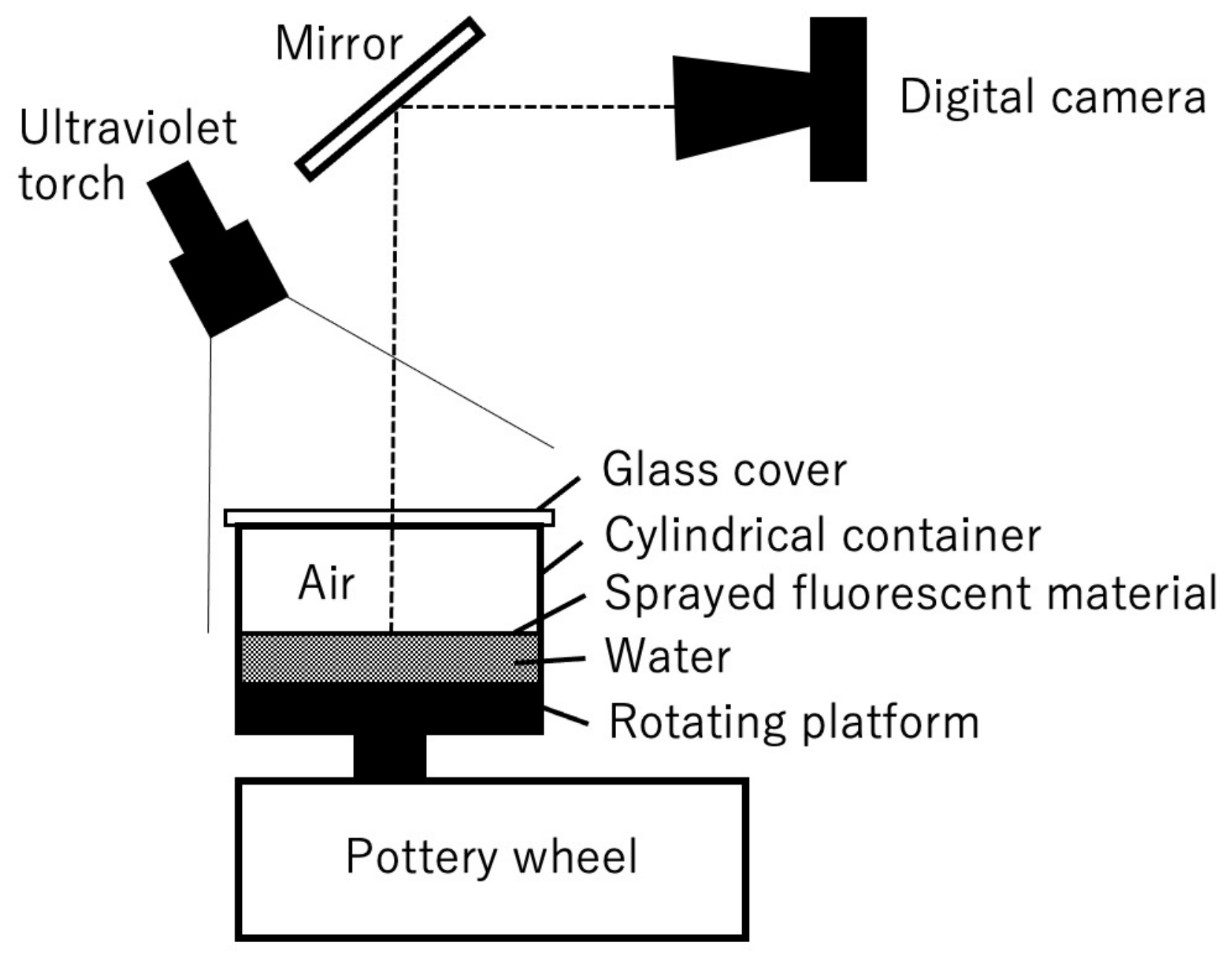}\caption{Experimental setup\label{fig1}}

\end{figure}
\begin{figure}

\includegraphics[scale=0.25]{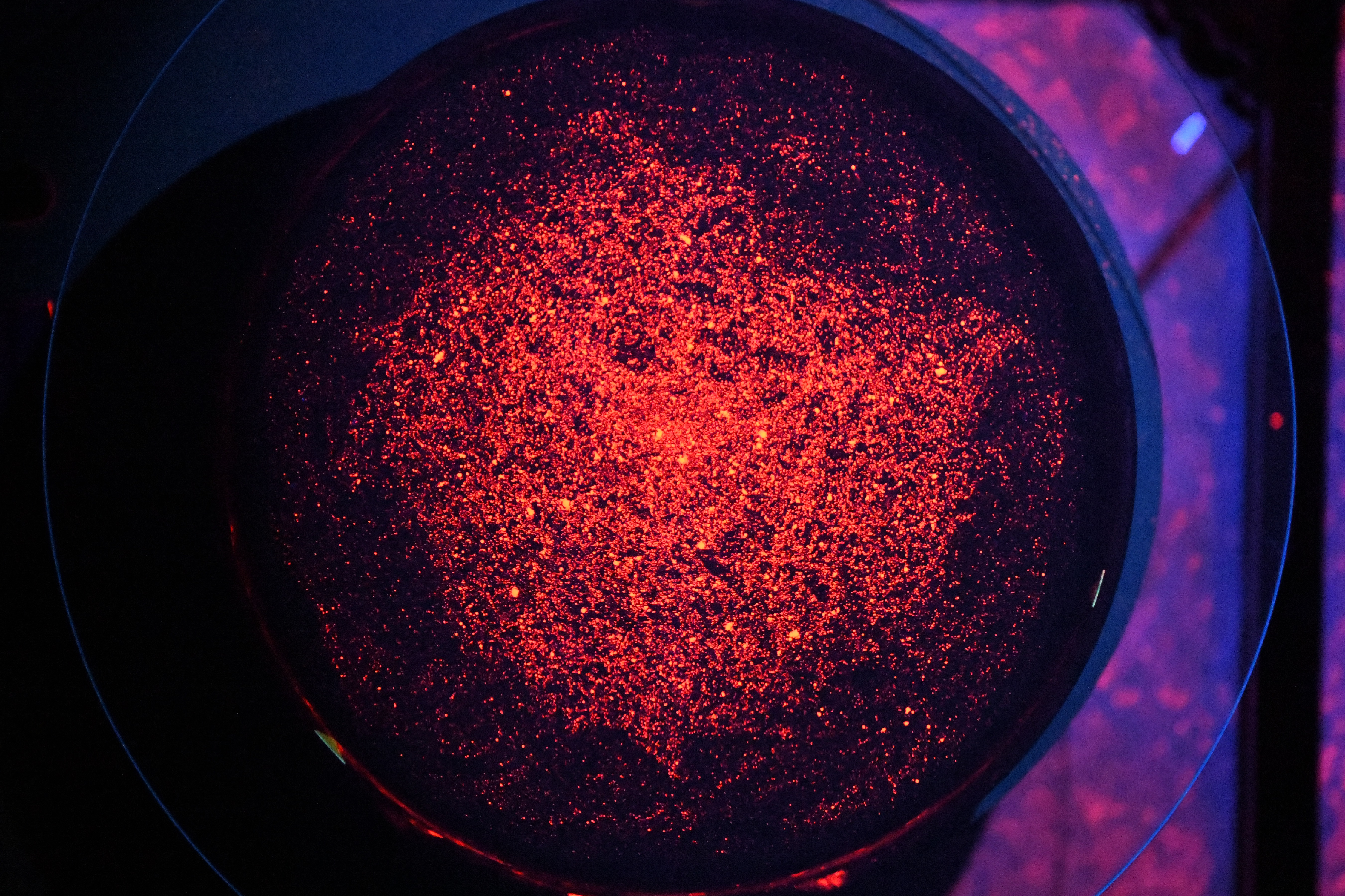}\caption{An example of a photo of a fluorescent top surface, illuminated by
an ultraviolet torch.\textcolor{red}{\label{fig_photo}} }

\end{figure}
The experimental setup of this study is shown in Fig. \ref{fig1}.
Water was poured by the depth of 8.0 cm into a horizontally-positioned,
acrylic and open cylindrical container with the inner diameter of
29 cm, and a solution (specific gravity: 0.94-0.95) of 11.6 wt\% fluorescent
material (Central Techno, Lumisis E-420) for ultraviolet laser-induced
fluorescence in predominantly oleic acid-based oil was sprayed one
time on the top water surface by a small atomizer so that the droplets
of the oil were distributed as uniformly as possible. The light of
peak wave length 613 nm is emitted by the material, excited by the
ultraviolet light of wave length between approximately 250 and 380
nm. The use of ultraviolet light prevents photos from halation, while
the visible light emission from the fluorescent material makes the
visualized flow clearer, suitable for low-speed PIV (Particle Image
Velocimetry).

The temperature of water was measured two times by a thermistor (Netsuken,
SN3000) before and after an experiment. The (absolute) difference
of the two temperatures exceeds 0.4 $^{\circ}\mathrm{C}$ for only
4 cases in total 67 experiments in this study, being within 0.1 $^{\circ}\mathrm{C}$
for 42 cases. The arithmetic mean of the temperatures was utilized
to evaluate the viscosity of water. The mean temperature at each experiment
ranges from 9.0 to 27.5 $^{\circ}\mathrm{C}$. As a result, the kinematic
viscosity of the water \citep{kestin1978,JISZ8803} changes from $8.465\times10^{-7}$
to $1.351\times10^{-6}\,\mathrm{m}^{2}/\mathrm{s}$, and just the
variation simulates the assessment with the exchange of low-viscosity
fluids. 

After the spray of the solution and the temperature measurement, a
glass cover that prevents inner water flow from the disturbance of
outer air flow in the laboratory was placed on the container. Although
an air layer of the depth of 3.0 cm ($\equiv D^{*'}$) is made between
the cover and the water surface, an accurate measurement of slow speed
on the surface was impossible without the cover.

After that, the container was horizontally rotated at a fixed speed
for ten minutes by an electric pottery wheel (Nidec-Shimpo, RK-3D),
and an axisymmetric flow was induced. The speed was also not fixed:
the maximum value of top-surface-mean speed ranges from $1.421\times10^{-3}$
to $2.057\times10^{-3}\,\mathrm{m}/\mathrm{s}$. After the cease of
the rotation, the top moving surface of the container, illuminated
by an ultraviolet torch (Central Techno, YKD-200, peak wave length:
365 nm) in a darkened room, was shot by a digital camera (Nikon, D850,
total pixels of the image sensor: 46.89 million) with an aspherical,
low-distortion lens (Nikon, AF-S NIKKOR 24-70mm f/2.8E ED VR) every
two seconds for an hour. Such a long sampling, for about 0.2 in dimensionless
time, was required because of the low kinematic viscosity of the water.
Each photo, an example is shown in Fig. \ref{fig_photo}, has 5408
by 3600 pixels, taken at the focal length of 105 mm. The time-lapse
movie, whose distortion is to be calibrated by a photo of a check
pattern, was processed by a PIV analyzer (Kato Koken, Flow Expert2D2C)
and the velocity field on the top surface was obtained. The time variation
of its mean speed was utilized to evaluate the decay rate of SDM.

\subsection{Standard and pseudo SDMs and classification scheme}

The evaluation of kinematic viscosity based on Eqs. (\ref{eq:lamda_SDM})
and (\ref{eq:evald_viscosity}) assumes that the open, top surface
of a cylindrical container is kept flat during an experiment. Since
the top surface is well approximated by the flat one after sufficiently
long time, it is certain that there exists the SDM described in Sec.
\ref{subsec:Fund_FMS}, hereafter referred to as the standard SDM. 

When a fluid speed is relatively large, however, the effect of a surface
wave, sloshing, is not negligible. An irrotational, (almost) inviscid,
linear wave theory deduces that the sloshing in a cylindrical container
has the SDM of a dimensionless decay rate $2x_{1}^{2}(\equiv\lambda_{SDM}^{(s)})$
with the distribution of azimuthal velocity component being the Bessel
function \citep{ibrahim2005}. The rate is independent of the aspect
ratio $\alpha$, caused by the linear-, infinitesimal-amplitude-wave
theory that neglects the effect of viscosity near wall surfaces. After
a long time, the effect has the fluid motion approached the standard
SDM asymptotically. The temporal, pseudo SDM is referred to as the
sloshing SDM. 

In order to measure the kinematic viscosity using such an open container,
it is crucial to properly classify each vanishing fluid mode into
one of the SDMs. The standard mode is observed when an initial surface
wave on the top surface, associated with initial fluid speed, is small
enough so that after a short period the distribution of azimuthal
velocity component is well approximated by the Bessel function. Such
time evolution is realized when the top surface is kept flat. Inversely,
the sloshing mode is observed when the initial speed is large enough
to actualize the initial wave. Note that physically high or low speed
depends on the kinematic viscosity of a test fluid. It is reasonable,
therefore, that we introduce a dimensionless criterion $U_{c}$ to
determine whether or not the initial speed, say the maximum dimensionless
top-surface-mean speed $U_{max}$, is high enough for producing initial
waves.

Once such a wave occurs, it takes much time to appear the Bessel-like
azimuthal distribution even if the initial speed is relatively low.
The experimental observation makes it virtually impossible to obtain
the decay rate $\lambda_{SDM}$ of the standard SDM within the accuracy
of velocimetry. Therefore, the condition that the azimuthal distribution
is well approximated by the Bessel function, i.e. the correlation
coefficient $c$ between the distribution and a correlation Bessel
function is larger than a criterion $c_{c}$, at a certain time $t_{c}$
after a long time must be a necessary condition for the standard SDM. 

Herein, we should note that the sloshing SDM is observed while the
surface wave is large enough. Eventually, such a wave is affected
by the viscous boundary layer in the vicinity of wall surfaces, and
the dimensionless decay rate is apart from $\lambda_{SDM}^{(s)}$.
When $U_{max}<U_{c}$, therefore, the fluid mode can not be regarded
as the sloshing SDM even if $c\leq c_{c}$.

Furthermore, we need a sufficiently long, straight-line, dimensionless
time range $T_{sl}$ on a semi-log plot of dimensionless relation
between top-surface-mean speed $U$ and time $t$ for evaluating the
decay rate accurately, regardless of whether or not a mode is the
standard SDM.

That is the reason why experimental vanishing fluid modes are classified
as follows. A mode is the standard SDM if

\begin{subequations}

\begin{equation}
U_{max}<U_{c},\,T_{sl}\geq T_{c},\;\mathrm{and}\;c>c_{c}\;\mathrm{at}\;t=t_{c},\label{eq:cond_stand}
\end{equation}
and the sloshing SDM if 

\begin{equation}
U_{max}\geq U_{c}\;\mathrm{and}\;T_{sl}\geq T_{c},\label{eq:cond_slosh}
\end{equation}
and otherwise discarded.\end{subequations}

\section{Results and Discussions}

The classification fully depends on the dimensionless critical values.
Firstly, the instant $t_{c}$ is set at $0.1$, i.e. $t_{c}^{*}=2103\,\mathrm{s}$
when a water temperature is 20 $^{\circ}\mathrm{C}$, so that the
time is more than twice as large as the decay time of $1/\lambda_{SDM}(=0.044)$
and at the instant the top-surface-mean speed $U$, estimated at $23.8(=\bar{U}_{max}\exp(-\lambda_{SDM}t_{c}))$,
is greater than the dimensionless speed-measurement limit of 10.7-17.1,
estimated from a typical dimensional limit of 0.1 mm/s, where in this
study the mean value $\bar{U}_{max}$ of $U_{max}$ is 232.0. In addition,
$c_{c}$ and $T_{c}$ are fixed at 0.995 and 0.06, respectively. The
pseudo, sloshing mode eventually agrees with the standard one, therefore
the criterion $c_{c}$ must be relatively higher value to distinguish
these modes with each other. The $T_{c}$ corresponds to 1262 s when
a water temperature is 20 $^{\circ}\mathrm{C}$. Although the accuracy
of the decay rate increases with $T_{c}$, the number of classified
SDMs decreases. In this study, the average of $T_{sl}$ for total
67 cases is 0.0737, and such a smaller critical value ensures a sufficient
number of SDMs. In contrast, the reasoning of the critical value $U_{c}$
is impossible. Therefore, the value must be treated as a parameter.
Taking $\bar{U}_{max}$ into consideration, therefore, we switched
the value between 223 and 240, and the effects on evaluated viscosity
are examined.

\begin{figure}
\includegraphics[scale=0.9]{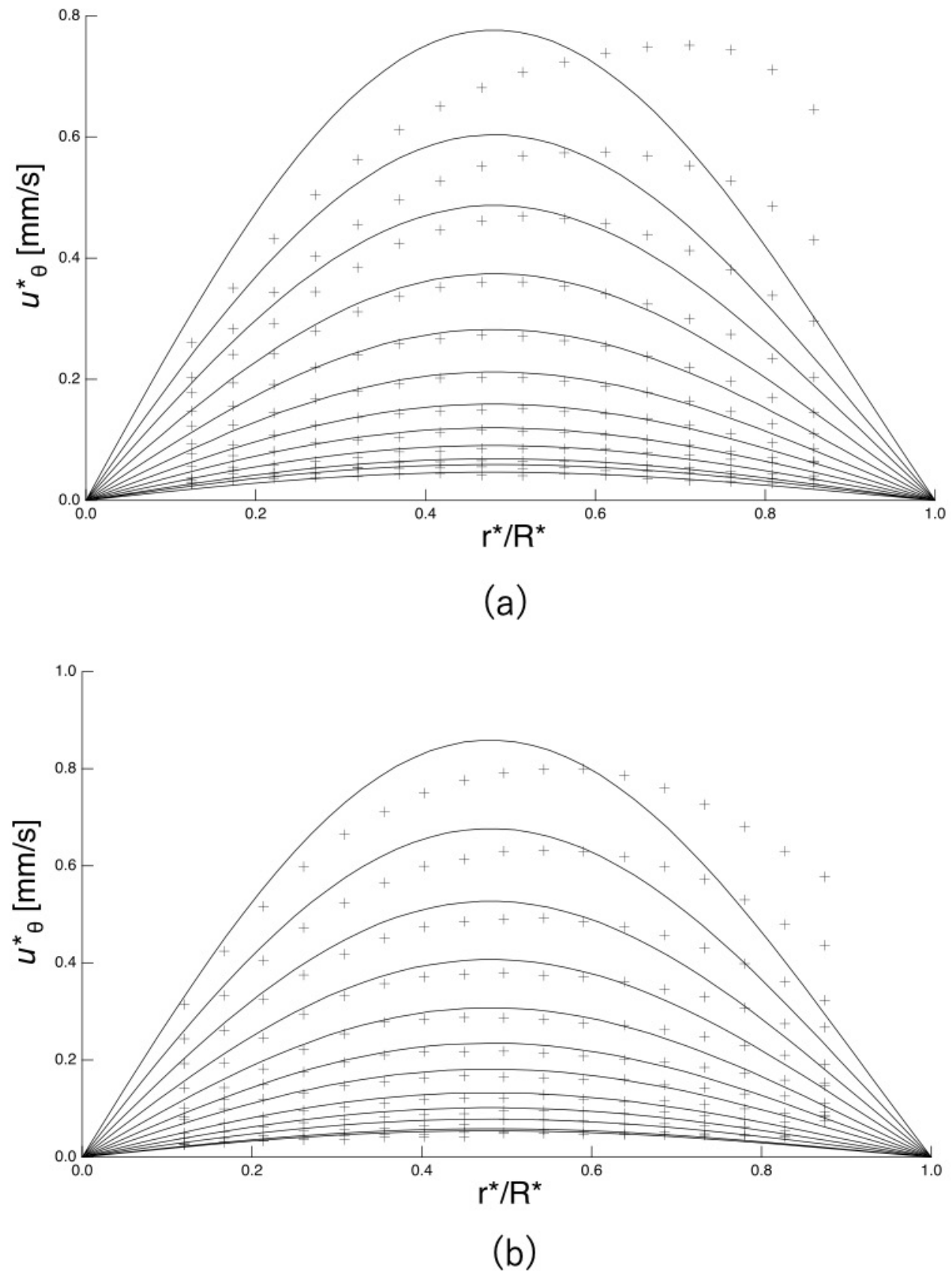}\caption{Typical distributions of azimuthal velocity component for (a) the
standard and (b) sloshing SDMs, drawn every 200 s from $t^{*}=1000$
to 3200 s. The measured distributions and their corresponding correlation
Bessel functions are indicated by markers and solid lines, respectively.
The uppermost is at $t^{*}=1000$ s. \label{fig2} }
\end{figure}
Typical distributions of azimuthal velocity component are shown in
Fig. \ref{fig2}. Fig. \ref{fig2}(a) shows an example (case A) of
the distribution of the standard SDM. In this case, the water temperature
is 12.2 $^{\circ}\mathrm{C}$, and $U_{max}$, $U_{max}^{*}$, $T_{sl}$
and $c$ at $t=0.1$ are $2.07\times10^{2}$, 1.76 mm/s, $8.80\times10^{-2}$
and 0.998, respectively. The distribution is well approximated by
the Bessel function with a slight difference near the side wall in
the sense that the correlation coefficient between the distribution
and the function exceeds 0.998 for $t^{*}\geq1800$ ($t>0.1$). On
the other hand, Fig. \ref{fig2}(b) is an example (case B) of the
distribution of the sloshing mode. The water temperature of the case
is 25.1 $^{\circ}\mathrm{C}$, and $U_{max}$, $U_{max}^{*}$, $T_{sl}$
and $c$ are $3.11\times10^{2}$, 1.91 mm/s, $8.07\times10^{-2}$
and 0.983, respectively. While the distribution is lower at the center,
it is higher near the side wall, when compared to correlation Bessel
functions. These figures show that $c_{c}$ must be over 0.99 for
apparent agreement with the Bessel function.

\begin{figure}
\includegraphics[scale=0.5]{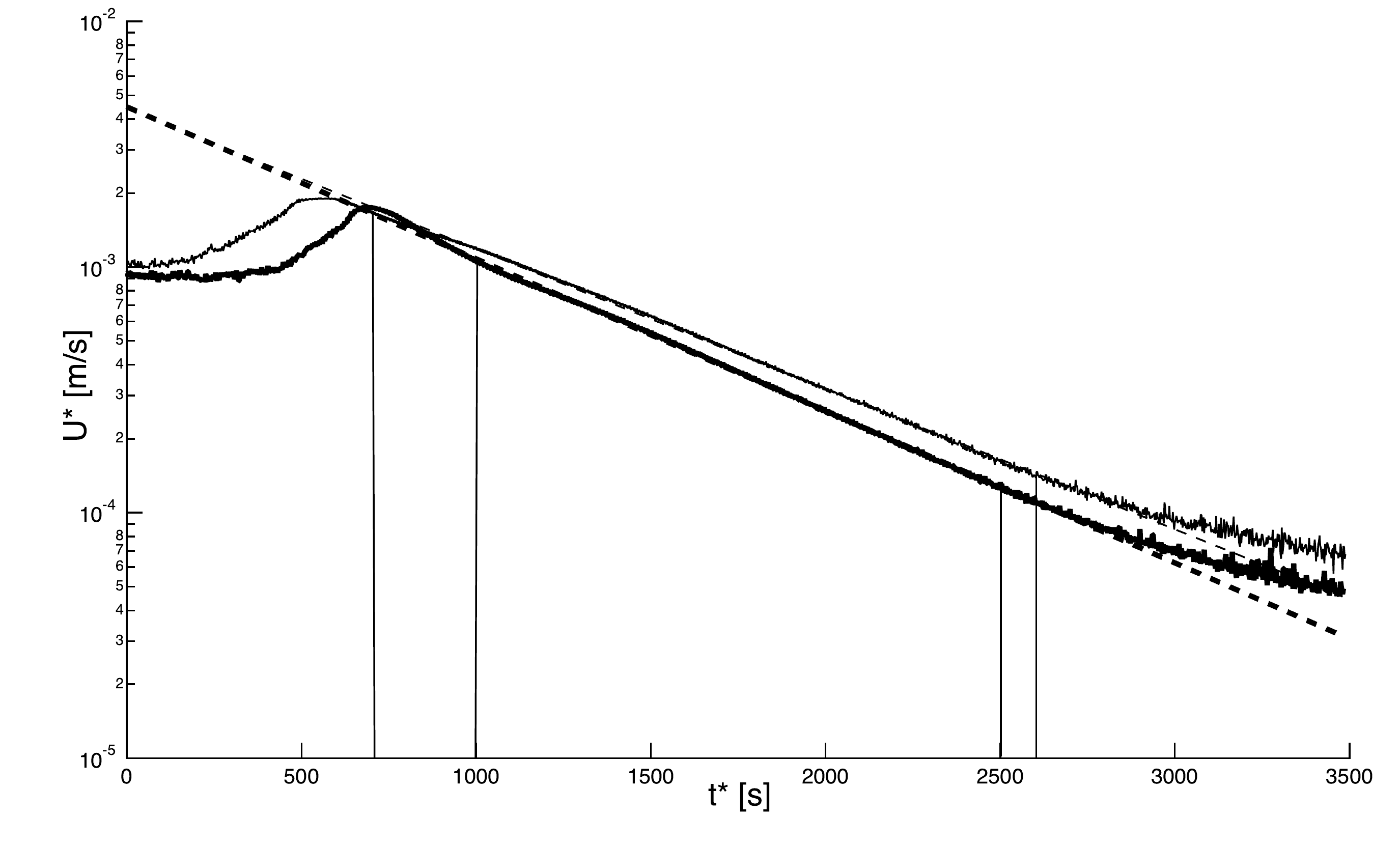}\caption{Time variation of top-surface-mean speed. Thick and thin solid lines
indicate the variation for cases A and B, respectively. Vertical lines
show the region of straight line on the semi-log plot, and two dashed
lines show the corresponding results of exponential fitting.\label{fig3} }
\end{figure}
The corresponding variations of top-surface-mean speed are shown in
Fig. \ref{fig3}. We can confirm that there exist linear regions between
corresponding two vertical lines on the semi-log plot and that each
region is well approximated by the corresponding exponential function.
It should be noted that the exponential decay begins even when the
distribution of azimuthal velocity component is far from the Bessel
function in comparison with Fig. \ref{fig2}. The property allows
us to evaluate the exponential decay rate accurately.

\begin{table}[t]
\caption{Average of normalized decay rates and their standard deviation for
each mode. The critical value $U_{c}$ is changed from 223 to 240.
\label{table1}}
\begin{tabular}{ccccc}
\hline 
$U_{c}$ & \multicolumn{2}{c}{223} & \multicolumn{2}{c}{240}\tabularnewline
slowest decaying mode (SDM) & standard & sloshing & standard & sloshing\tabularnewline
\hline 
data number & 6 & 21 & 10 & 13\tabularnewline
analytical decay rate $(\lambda_{SDM}\;\mathrm{or}\;\lambda_{SDM}^{(s)})$ & 22.8 & 29.4 & 22.8 & 29.4\tabularnewline
mean decay rate $\bar{\lambda}$ & 23.9 & 29.5 & 23.7 & 32.1\tabularnewline
unbiased variance $V$ & 3.68 & 32.3 & 7.39 & 24.2\tabularnewline
standard deviation $\sigma(=\sqrt{V})$ & 1.92 & 5.69 & 2.72 & 4.92\tabularnewline
relative error to analytical value $(=\sigma/\lambda_{SDM}^{(\bullet)})$ & 0.084 & 0.194 & 0.119 & 0.168\tabularnewline
\hline 
\end{tabular}
\end{table}
Similarly, the exponential decay rates for the other cases are evaluated
and normalized by the kinematic viscosity at each water temperature.
The total result of uncertainty analyses is shown in Table \ref{table1}. 

Firstly, it is found that the theoretical decay rates $\lambda_{SDM}$
and $\lambda_{SDM}^{(s)}$ are included in the average value $\bar{\lambda}$
of decay rates $\pm$ its standard deviation $\sigma$ for each SDM,
i.e. $\left|\bar{\lambda}-\lambda_{SDM}\right|<\sigma$, indicating
that the classification of the two modes is appropriate. Under the
condition that the relation holds and that $\nu$ is constant, the
relative error of measured kinematic viscosity $\nu_{m}$, estimated
by Eq. (\ref{eq:evald_viscosity}), to the real value $\nu$ is bounded
above by that of dimensionless decay rate $\lambda$ to $\lambda_{SDM}$
because of the following inequality

\begin{equation}
\left|\frac{\bar{\lambda}}{\lambda_{SDM}}-1\right|=\left|\frac{\bar{\lambda^{*}}L^{2}/\nu}{\bar{\lambda^{*}}L^{2}/\bar{\nu}_{m}}-1\right|=\left|\frac{\bar{\nu}_{m}}{\nu}-1\right|\leq\frac{\sigma}{\lambda_{SDM}}.\label{eq:relative_error}
\end{equation}
It follows that the relative error of measured kinematic viscosity
with the temperature of a test fluid fixed by a thermostat under a
constant ambient pressure is less than or equal to the relative error
shown in Table \ref{table1}. Recall that dimensionless properties
obtained by experiments are independent of a kind of fluid.

The deviation $\sigma$, i.e. error, of dimensionless decay rate for
the standard SDM, is always smaller than that for the sloshing SDM,
independent of $U_{c}$. It follows that relatively accurate measurement
can be achieved by use of the standard SDM. 

We can also find that the error for the standard SDM increases with
$U_{c}$ and that more accurate measurement can be conducted by diminishing
an initial speed. In order to take a longer $T_{sl}$, however, more
accurate velocimetry should be utilized. In Fig. \ref{fig3}, the
measured speed is apart from the exponential correlation function
and begins to fluctuate when the speed is below 0.1 mm/s ($\equiv u_{lim}^{*}$),
at which it reaches the limit of measurement. The condition (\ref{eq:cond_stand})
for the standard SDM leads to

\[
U_{c}e^{-\lambda_{SDM}T_{c}}>U_{max}e^{-\lambda_{SDM}T_{sl}}>u_{lim}.
\]
In order to ensure a sufficient data number of the standard SDM, $u_{lim}/U_{c}(=u_{lim}^{*}/U_{c}^{*})$
must be small enough when compared to $\exp(-\lambda_{SDM}T_{c})$.
If we can reduce $u_{lim}$ to half, we can also reduce $U_{c}$ to
half with $T_{c}$ fixed, and it contributes to more accurate measurement
by the standard SDM. 

In contrast, the error for the sloshing SDM decreases with $U_{c}$.
However, it is insufficient for accurate measurements. As long as
an initial speed is large enough, the sloshing SDM occurs frequently
and makes it possible for us to measure easily the kinematic viscosity
with lower accuracy. For example, the method can be utilized for a
device to notify the necessity of the exchange of machine oil.

Finally, it is proper to point out that the increase of the mean decay
rate up to 0.2$-$0.4 for the standard mode can be explained by the
glass cover. In fact, axisymmetric, radial-component-free, two-dimensional
simulations in a closed cylindrical container with the air and water
layers coupled, their interface kept flat, and the depth ratio $D^{*'}/D^{*}$
fixed at 3/8 show that the dimensionless decay rate of the SDM changes
from 23.0 ($5^{\circ}\mathrm{C}$) to 23.2 ($30^{\circ}\mathrm{C}$).
Although the details are omitted in this paper, the ratio of the dynamic
viscosity of air to that of water, i.e. $1.15\times10^{-2}-2.35\times10^{-2}$
for $5-30^{\circ}\mathrm{C}$, convinces us of the rate increase.
The deviation about 0.3 from the analytical value 22.8 and its dispersion
$\pm0.1$ are far smaller than the standard deviation $\sigma$, which
is greater than 1.92. Similarly, the effect of the cover on the decay
rate of the sloshing SDM is expected to be the order of 1\%. The advantage
of the cover in the reduction of $\sigma$ surpasses the disadvantage
in this study.

\section{Concluding Remarks}

This study presents a new method to measure accurately the kinematic
viscosity of fluids by use of a spectroscopy decomposing a stirred,
vanishing velocity field into fluid modes (Stokes eigenmodes). The
method, Fluid Mode Spectroscopy (FMS), is based on the fact that each
Stokes eigenmode has its inherent decay rate of eigenvalue and that
the dimensionless rate of the slowest decaying mode (SDM) is constant,
dependent only on the normalized shape of a fluid container. The decay
rate is obtained analytically for some shapes like an open cylindrical
container of this study. The FMS supplements major conventional measuring
methods with each other, particularly useful for measuring low kinematic
viscosity. The main results are as follows:
\begin{enumerate}
\item In order to avoid numerical errors for evaluating $\lambda_{SDM}$
and to use easier flow visualization, an open cylindrical container
of this study was the best. However, an open container involves a
pseudo, sloshing decay mode. Therefore, we have no choice but to classify
each vanishing fluid mode into the mode and a viscous, surface-flat
standard mode.
\item The classification depends on four dimensionless criteria, i.e. the
maximum velocity $U_{c}$, linearly time range $T_{sl}$ on semi-log
plot of the relation between the surface-averaged speed and time,
and correlation coefficient $c_{c}$ between a distribution of azimuthal
velocity component and its correlation Bessel function at an instant
$t_{c}$. 
\item By use of the standard mode in an open cylindrical container the kinematic
viscosity of water is measured within a relative error smaller than
8.4-11.9\%. Taking that water is a typical example of low-viscosity
fluids into consideration, the measurement is accurate. The smaller
$U_{c}$, more accurate the measurement becomes. Since the value is
proportional to a lower limit of velocity measurement, the accuracy
fully depends on the used velocimetry.
\item By use of the sloshing mode, the kinematic viscosity is measured within
a relative error smaller than 16.8-19.4\%. The mode is frequently
observed when an initial speed is high enough, useful for easier,
low-precision measurements, such as a device to notify the necessity
of the exchange of machine oil.
\item Thus, within the error of each vanishing fluid mode, the applicability
and methodology of FMS are validated.
\end{enumerate}
In principle, the method is applicable to any type of fluid whose
visualization is possible. In order to increase initial speed and
S/N ratio thereby, the measurement must be ultimately classification-free.
It virtually makes the critical value $U_{c}$ infinity, and we can
take a sufficiently large $U_{max}$ for an experiment based on the
standard SDM. There are two methods for the purpose. One is the method
to utilize a closed container. For example, we can analytically obtain
$\lambda_{SDM}$ even if a cylindrical container is closed, as described
in the next section. The rate allows us to apply FMS to closed cylindrical
containers. The other is the method to equate the decay rate $\lambda_{SDM}$
of the standard SDM with that of the sloshing mode, $\lambda_{SDM}^{(s)}$.
It is achieved when the aspect ratio $\alpha$ of the fluid layer
in a cylindrical container is equal to $\pi/(2x_{1})$. 

The effects of other parameters on the accuracy of FMS are of interest.
For example, the mere asperity, roughness, of a container is expected
not to affect the accuracy, because the thickness of a velocity boundary
layer on a wall surface reaches the order of the radius or depth of
the container after a long time. However, if the wall is made by a
porous media, its macroscopic skin friction may be quantified by the
decay rate of FMS. They are issues in the future.
\begin{acknowledgments}
HI is grateful to M. Miyahara and M. Suzuki of Central Techno corporation
for providing the technical information of the solution of fluorescent
material in oil. HI is also grateful to Dr. Y. Ueda of Setsunan University
for enlightening discussions on the possible modes in open cylindrical
containers.
\end{acknowledgments}

\section*{Author Declarations}

\subsection*{Intellectual property}

HI, NI, TH, and AK have a pending patent (publication number: JP,
2021-063675, A) for FMS. MH has a licensed Japanese patent (No. 6713598)
for the fluorescent material utilized in this experimental study.

\section*{Data Availability}

The data that supports the findings of this study are available within
the article.

\appendix

\section{Axisymmetric, radius-component-free Stokes eigenmodes in cylindrical
containers\label{sec:Stokes-modes-cc}}

Now let us consider axisymmetric Stokes eigenmodes in horizontally
positioned cylindrical containers with flat top and bottom faces.
Particularly, we shall confine to the discussion for the case that
the radius component $u_{r}$ of velocity is zero in this section.
The condition allows us to decouple the pressure term from the Navier-Stokes
equation system on the cylindrical coordinates. The dimensionless
eigen equation for $u_{\theta}$, normalized by the kinematic viscosity
$\nu$ and the inner radius $R^{*}$, is of the form

\[
-\lambda u_{\theta}=\frac{\partial}{\partial r}\left(\frac{\partial u_{\theta}}{\partial r}+\frac{u_{\theta}}{r}\right)+\frac{\partial^{2}u_{\theta}}{\partial z^{2}},
\]
and it is straightforward to solve the equation by the method of separation
of variables. 

If the flat, top surface is open, free-slip condition $\partial u_{\theta}/\partial z=0$
is subjected to the top surface, while no-slip condition, $u_{\theta}=0$,
is exerted on the bottom ($z=0$) and side ($r=1$) walls and on the
axis ($r=0$). Then, the Stokes eigenmodes and their corresponding
eigenvalues are found to be

\begin{subequations}

\[
u_{\theta(n,m)}\sim J_{1}(x_{n}r)\sin\left(\frac{2m+1}{2\alpha}\pi z\right),
\]
and

\[
\lambda_{n,m}=x_{n}^{2}+\left(\frac{2m+1}{2\alpha}\pi\right)^{2},
\]
\end{subequations}respectively, where the solution is labeled by
a positive integer $n$ and a nonnegative integer $m$, and a relation
$A\sim B$ indicates that $A$ is proportional to $B$. $J_{1}$ is
the Bessel function of the first kind of order unity, $x_{n}$ ($n=1,2,\cdots)$
positive $n$th zero of the function. It follows that the slowest
decaying mode and its corresponding eigenvalue can be expressed as
Eqs. (\ref{eq:Bessel_dist}) and (\ref{eq:lamda_SDM}), respectively.

On the other hand, if all faces are closed, i.e. if no-slip condition
is also exerted on the top face, then the solution can be expressed
as

\begin{subequations}

\[
u_{\theta(n,m)}\sim J_{1}(x_{n}r)\sin\left(\frac{m\pi z}{\alpha}\right),
\]
and

\[
\lambda_{n,m}=x_{n}^{2}+\left(\frac{m\pi}{\alpha}\right)^{2},
\]
\end{subequations}where integers $n,m$ are positive. In this case,
therefore, the smallest decay rate $\lambda_{SDM}=x_{1}^{2}+(\pi/\alpha)^{2}.$

\section{Numerical simulations in cylindrical containers with a flat, open
top surface\label{sec:simulation}}

In this study numerical simulations were also conducted in order to
validate the standard SDM of the eigenmode (\ref{eq:Bessel_dist})
with the eigenvalue (\ref{eq:lamda_SDM}), and to examine properties
when the top free-surface is always flat. Dimensionless and incompressible
Navier-Stokes equation system without forcing on the cylindrical coordinates
is discretized by finite volume method.

No-slip boundary condition, $\mathbf{v}=\mathbf{0}$, is subjected
to the side ($r=1$) and bottom ($z=0$) wall surfaces, and periodic
condition is exerted on the faces at $\theta=0$ and $2\pi$. The
other faces are subjected to free-slip conditions in the sense that

\[
\frac{\partial u_{r}}{\partial z}=\frac{\partial u_{\theta}}{\partial z}=u_{z}=0\;\mathrm{at}\;z=\alpha,
\]
and that

\[
u_{r}=u_{\theta}=\frac{\partial u_{z}}{\partial r}=0\;\mathrm{at}\;r=0.
\]

Convection and diffusion terms are discretized by QUICK and central
difference schemes, respectively. The differenced equation system
is timely evolved by explicit SMAC method. A division number $N$
in the radius ($0\leq r\leq1$), azimuth ($0\leq\theta\leq2\pi$)
and height ($0\leq z\leq\alpha$) directions is held fixed at 30 except
for the case of accuracy assessment. 
\begin{figure}
\includegraphics[scale=0.5]{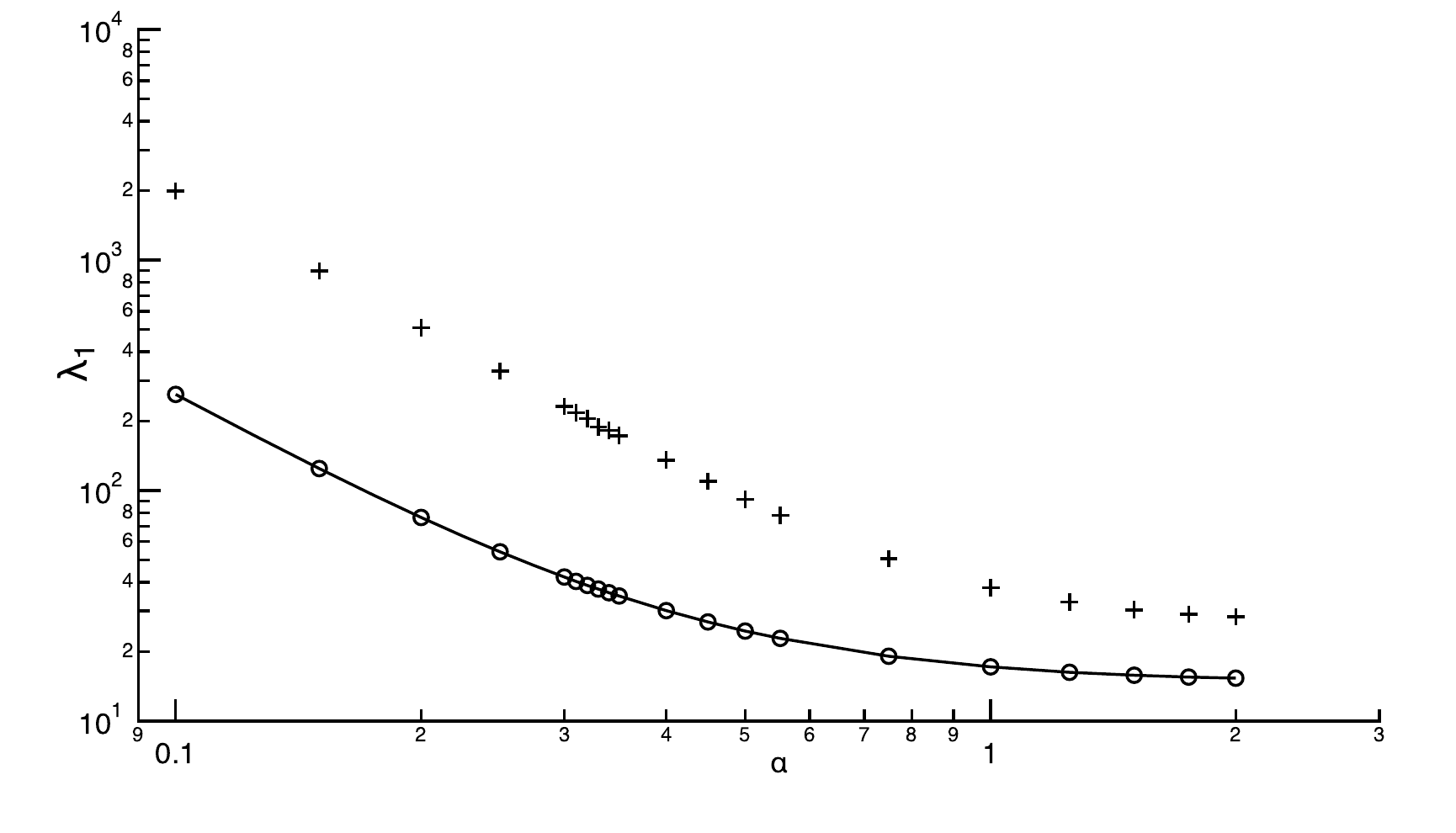}\caption{Variation of the smallest eigenvalue $\lambda_{1}$ with respect to
$\alpha$. Solid line is analytical eigenvalue (\ref{eq:lamda_SDM}).
Circle and cross indicate eigenvalues for the cases of ``$u_{r}=0$''
and ``$u_{\theta}=0$'', respectively. The eigenvalue of the $u_{r}$-free
mode, less than that of the $u_{\theta}$-free mode, agrees with that
of the standard mode, Eq. (\ref{eq:lamda_SDM}), and the corresponding
distribution, omitted due to space constraints, also agrees with that
of the standard mode (\ref{eq:Bessel_dist}). \label{figA}}
\end{figure}

Firstly, let us consider the problem of whether the standard SDM (\ref{eq:Bessel_dist})
is the true SDM. Numerical eigenvalue analyses of the Stokes eigenmodes
with the aid of the above-mentioned numerical integration were performed
for two axisymmetric initial velocity fields as follows: 

\begin{eqnarray*}
\mathrm{(case\,"}u_{r}=0\mathrm{")} & \;\mathbf{v}_{0} & \equiv(\left.u_{r}\right|_{t=0},\left.u_{\theta}\right|_{t=0},\left.u_{z}\right|_{t=0})=(0,4ar(1-r)z^{2},0),\\
\mathrm{(case\,"}u_{\theta}=0\mathrm{")} & \;\mathbf{v}_{0} & =(-a\sin(2\pi z)\sin(\pi r),0,a\sin(2\pi r)\sin(\pi z)).
\end{eqnarray*}

The SDM and its corresponding eigenvalue were computed by adjusting
the norm of a vector field to unity at each time step. The steady
vector field and its corresponding negative increasing rate of the
norm before the adjustment at each step agree with the SDM and its
corresponding eigenvalue, respectively. Note that, therefore, the
magnitude $a$ does not affect the result. It is well known that eigenvector
fields of vorticity for the Stokes eigenmode are normal to each other,
and therefore we must have different SDMs for the two cases.

The variation of the smallest eigenvalue with respect to the aspect
ratio $\alpha$ of the fluid layer are shown in Fig. \ref{figA}.
As expected, two different SDMs are obtained for each initial velocity.
However, the eigenvalue of the case ``$u_{\theta}=0$'' is always
larger than that of the $u_{r}$-free, standard mode, Eq. (\ref{eq:lamda_SDM}).
Particularly, for the case $\alpha\leq0.32$, the eigenmode for the
case ``$u_{\theta}=0$'' is transient, remaining for a finite period
of time and eventually shifting to the other SDM. That is to say,
the $u_{\theta}$-free mode is linearly unstable, and the transition
is caused by round-off errors involved in numerical simulations. These
results indicate that the standard SDM is the true SDM for $0.1\leq\alpha\leq2$,
computed in this study. 

Similar to the $u_{r}$-free mode, $\lambda_{1}$ of the $u_{\theta}$-free
mode behaves like $\sim\alpha^{c}$ when $\alpha$ is small enough
with a power low exponent $c$ being slightly smaller than that of
the $u_{r}$-free mode, and saturates for larger values. It implies
that the standard SDM is the true one for any positive $\alpha$.
We can also ascertain that the slowest decaying mode is really axisymmetric,
$u_{r}$-free one (\ref{eq:Bessel_dist}) even for a general, non-axisymmetric
case of the form

\[
\mathbf{v}_{0}=(-a\sin(2\pi z)\sin(\pi r),4ar(1-r)z^{2}+a\sin(b\theta),a\sin(2\pi r)\sin(\pi z)),
\]
and they are the pieces of numerical evidence that the standard SDM
(\ref{eq:Bessel_dist}) is the true SDM. 

Next, common numerical integration without norm adjustment was conducted
with an initial axisymmetric velocity field of the form $\mathbf{v}_{0}=(0,r\omega,0)$
except for the no-slip surfaces, where $\omega=348$ so that the initial
top-surface-mean speed would be 232. The aspect ratio $\alpha$ was
held fixed at 0.5517. Such an initial velocity distribution, rigid-body
rotation, mimics typical experimental one of this study. 

\begin{figure}
\includegraphics[scale=0.5]{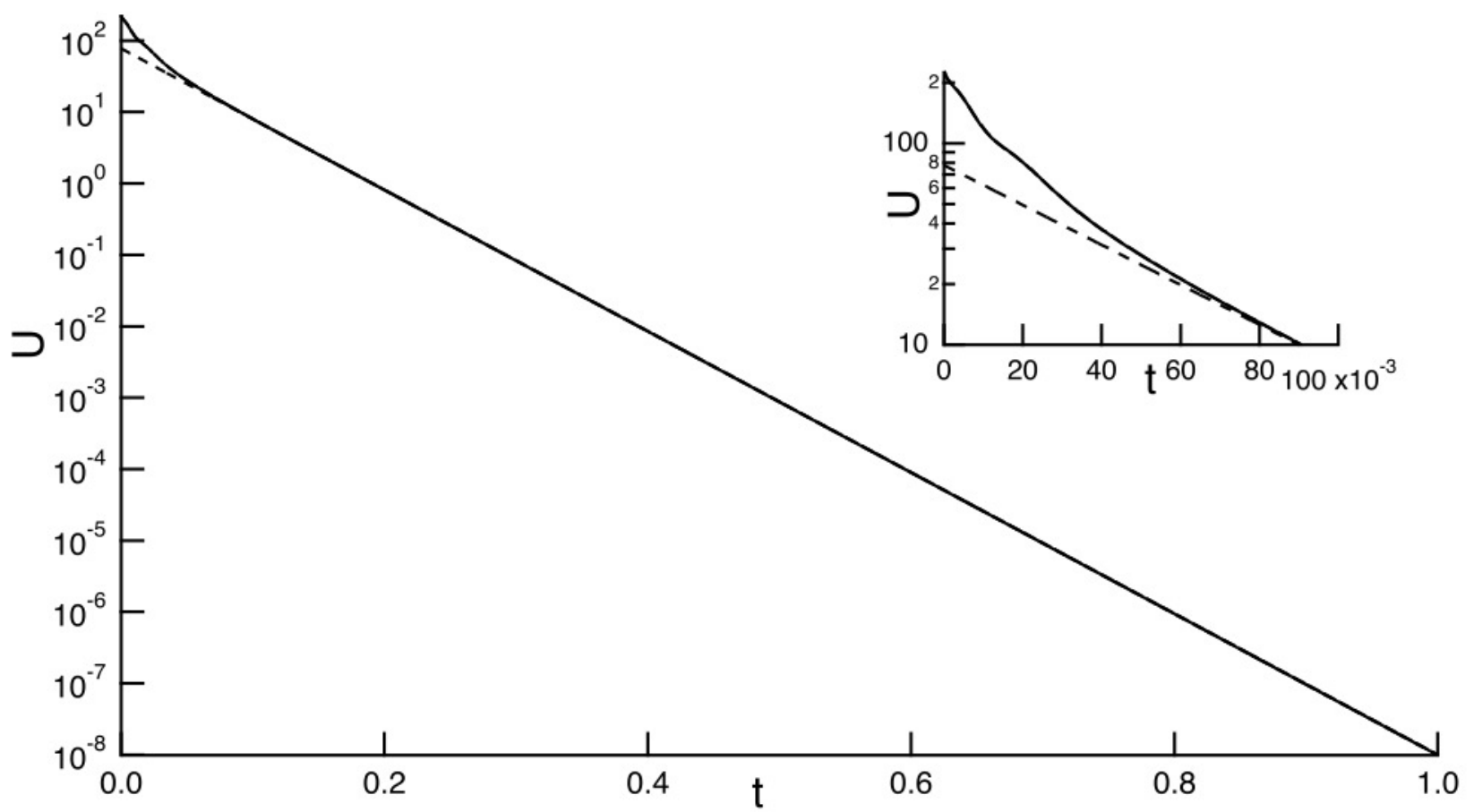}\caption{Variation of top-surface-mean speed with respect to time $t$. The
magnification is at around $t=0$. Dotted and solid lines indicate
the variations for $N=20$ and 30, respectively. But they are overlapped.
For $t>0.059$ the solid line agrees well with its exponential correlation
function, indicated by dashed line, and recovers the decay rate of
22.788(=$\lambda_{SDM}$) with the relative error of 0.0092\%.\label{figB}}
\end{figure}

\begin{figure}

\includegraphics[scale=0.4]{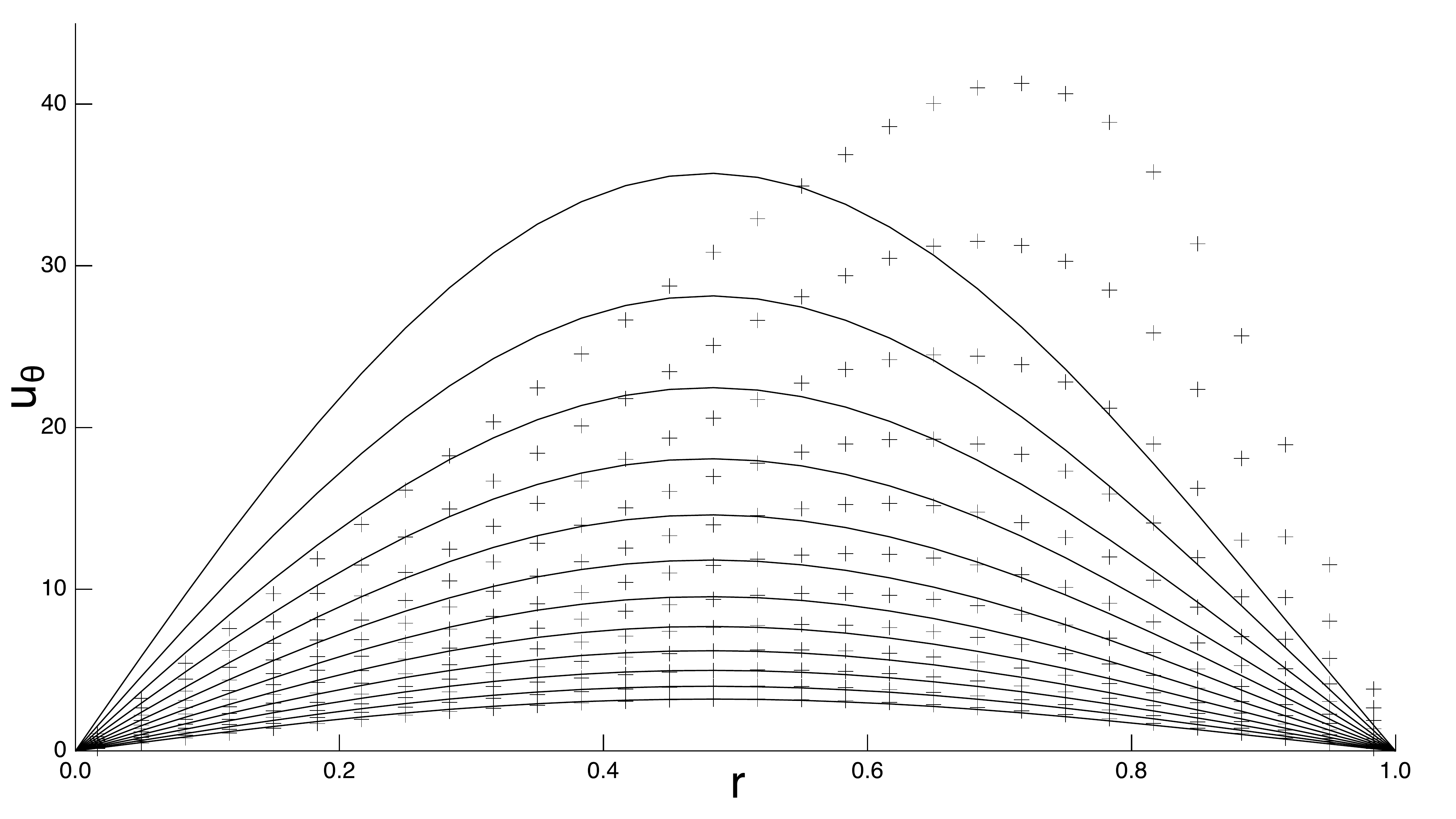}\caption{Decays of the distribution of azimuthal velocity component from $t=0.05$
to 0.16 ($N=30$). The distribution and its corresponding correlation
Bessel function are indicated by plus marker and solid line, respectively.
The uppermost is at $t=0.05$, and time increments are 0.01. The lowermost
distribution agrees well with the Bessel function: the correlation
coefficient exceeds 0.995 at $t=0.16$. \label{figC}}

\end{figure}

The time variations of top-surface-mean speed $U$ and that of the
distribution of azimuthal velocity component for $N=30$ are shown
in Figs. \ref{figB} and \ref{figC}, respectively. They show that
the exponential decay with the rate of $\lambda_{SDM}$ begins at
around $t=0.059$, even when the distribution of azimuthal velocity
component is far from the Bessel function. The onset of the Bessel-like
azimuthal distribution appears at $t=0.1451$, where the correlation
coefficient $c$ is 0.99. The coefficient is raised up to 0.995 at
$t=0.1579$. 

In this study the instant $t_{c}$ at which a decaying mode is classified
is set at 0.1, less than 0.1579. But the above-mentioned results depend
on an initial velocity distribution $\mathbf{v}_{0}$. For example,
for $\omega=25$ we can confirm that $c$ reaches 0.995 at $t=0.0814$.
Actual initial distributions are far from the rigid-body rotation,
dependent on a water temperature and a rotational number of each experiment.
Moreover, note that the maximum speed at $t=0.16$, shown in Fig.
\ref{figC}, is much smaller than 10.7-17.1, estimated from a typical
dimensional measurement limit of 0.1 mm/s. In contrast, the top speed
is maintained at around the limit at $t=0.1$. It is reasonable, therefore,
that $t_{c}$ is fixed at 0.1 in this study. 

Although the presentation of the results is omitted due to space constraints,
it is easy to perform computations with more different initial conditions
for the same aspect ratio $\alpha$. The results show that the exponential
decay begins at $t\simeq0.06$, independent of an initial velocity
distribution $\mathbf{v}_{0}$, because higher modes decays within
$t\leq1/\lambda_{SDM}=0.044$. 

\bibliographystyle{aipnum4-2}
\bibliography{fms}

\end{document}